\documentclass[reqno]{amsart} 
\usepackage{amsmath}%
\usepackage{amsfonts}%
\usepackage{amssymb}%
\usepackage{graphicx}

\usepackage{titlesec} 
\renewcommand\thesection{\Roman{section}} 
\renewcommand\thesubsection{\roman{subsection}} 
\titleformat{\section}[block]{\large\scshape}{\thesection.}{1em}{} 
\titleformat{\subsection}[block]{\large}{\thesubsection.}{1em}{} 

\usepackage[sc]{mathpazo} 
\usepackage[T1]{fontenc} 
\usepackage{microtype} 
\usepackage[hang, small,labelfont=bf,up,textfont=it,up]{caption} 
\usepackage{booktabs} 
\usepackage[english]{babel} 
\usepackage{titling} 
\usepackage{hyperref} 
\usepackage{array}
\usepackage{multirow}
\usepackage{hyperref} 
\usepackage{makecell}
\usepackage{setspace}
\usepackage[hmarginratio=1:1,top=32mm,columnsep=20pt]{geometry} 
\usepackage{authblk}

\usepackage{fancyhdr} 
\usepackage{abstract} 
\usepackage{titlesec} 
\renewcommand\thesection{\Roman{section}} 
\renewcommand\thesubsection{\roman{subsection}} 
\titleformat{\section}[block]{\large\scshape}{\thesection.}{1em}{} 
\titleformat{\subsection}[block]{\large}{\thesubsection.}{1em}{} 

\usepackage{abstract} 
\pretitle{\begin{center}\Huge\bfseries} 
\posttitle{\end{center}} 
\title{Statistical Modeling of Atmospheric Mean Temperature in sub$-$Sahel West Africa} 

\author[*]{Falaiye O. A.}
\author[*]{Sukam Y. M.}
\author[**]{Abimbola O. J.}
\affil[*]{Department of Physics, Faculty of Physical Sciences, University of Ilorin, Nigeria}
\affil[**]{Department of Physics, Faculty of Science, Federal University Lafia, Nigeria}

\date{} 
\renewcommand{\maketitlehookd}{%
\begin{abstract}
Atmospheric mean temperature $T_m$, is a vital parameter in the evaluation of precipitable water vapor (PWV) through the analysis of GPS signal, it is therefore important to have a good way of evaluation of Tm for the eventual accurate evaluation of PWV using GPS. Simple statistical models exist for various regions of the world for the evaluation of $T_m$  using surface temperature $T_s,$ in the form $T_m=aT_s+b$ where $a$ and $b$ are constants. For West Africa, where atmospheric data is usually very scarce, there is a minimal attempt at finding a statistical model for $T_m$, as a function of $T_s$. In this work attempt has been made to find such a model using data from the Climate Monitoring Satellite Facilities (CM-SAF) of the European Meteorological Satellites (EUMETSAT). The model derived was found to compare well with that obtained using radiosonde data with root-mean-square error of $1.189$ and mean-biased error of $0.0952$ between the two models.\newline
\textbf{Keywords:} West Africa, atmospheric mean temperature, precipitable water vapor.
\end{abstract}
}

\begin{document}

\maketitle

\section{Introduction}

Preciptable water vapor (PWV) is an important atmospheric parameter essential in both weather and climatic prediction. Knowledge of the variability of PWV is also very important in astronomy as water vapor interact with the incoming electromagnetic waves in the atmosphere.

GPS meteorology offers a real-time continuous measurement of PWV. From the GPS data, it is possible to estimate the zenith total delay (ZTD). The zenith hydrostatic delay (ZHD), which is the delay error caused by the dry component of the atmosphere could be estimated using models such as the Saastamoinen \cite{Saastamoinen:1972dg} model. The delay error introduced by the water vapor in the atmosphere, called the zenith wet delay (ZWD) is calculated thus:

\begin{equation}
\label{eq:1}
ZWD=ZTD-ZHD
\end{equation}

The ZWD is usually transformed into the PWV using the transformation equation as given below

\begin{equation}
\label{eq:2}
PWV=\Pi \times ZWD
\end{equation}

The transformation constant $\Pi$, is given by

\begin{equation}
\label{eq:3}
\Pi = \left[ 10^{-6} \left( \frac{k_{3}}{T_{m}} + k^{'}_{2} \right) \rho_{v} R_{v} \right] ^{-1}
\end{equation}

here, $\rho_{v}$ is the liquid water density = $1000\ kgm^{-3}$, $R_{v}$ is water vapor gas constant = $461.524\ JK^{-1}kg^{-1}$, $k_{3} = 377600\ K^{2}hPa$, $k^{'}_{2} = 22.1\ KhPa^{-1}$  and $T_m$ is called weighted atmospheric mean temperature determination of which is very crucial to accurate transformation of ZWD into the PWV.

The weighted mean atmospheric temperature $T_m$, could be defined as

\begin{equation}
\label{eq:4}
T_m=\frac{\int^{p_1}_{p_2}Td(lnp)}{\int^{p_1}_{p_2}d(lnp)}=\frac{\int^{\infty}_{z}\frac{e}{T}dz}{\int^{\infty}_{z}\frac{e}{T^2}dz}
\end{equation}

where $p_1$ and $p_2$ are pressures at two different layers of the atmosphere. $T$ is the ambient temperature. $e$ is the vapor pressure while $z$ is the height above the ground. From Eq.$\;\ref{eq:4}$. \;$T_m$ could be estimated from the radiosonde data taken at different layers of the atmosphere. However, such radiosonde data are usually not readily available across the globe, especially Africa, hence several attempts have been made to statistically relate $T_m$ with the surface temperature $T_s$ which is readily available as a basic meteorological parameter at any weather observing station. Bevis \textit{et al.}\cite{Bevis:1992dg} used $8700$ radiosonde profiles from $13$ stations in the United States of America (USA) to obtain a statistical model of the form $T_m=aT_s+b$. The Bevis \textit{et al.}\cite{Bevis:1992dg} model was adopted in many studies (e.g., Raju \textit{et al.}\cite{Raju:2007dg}; Fernandez \textit{et al.}\cite{Fernandez:2010dg}; Musa \textit{et al.}\cite{Musa:2011dg}; Abimbola \textit{et al.}\cite{Abimbola:2017dg}) across the globe despite the fact that it was obtained for the USA region. Mendes \textit{et al.}\cite{Mendes:2000dg} and Solbrig\cite{Solbrig:2000dg}, using radiosonde data, obtained similar linear relation for Germany, Raju \textit{et al.}\cite{Raju:2007dg} for Indian sub-continent, Shoji\cite{Shoji:2010dg} for Japan and Isioye \textit{et al.}\cite{Isioye:2016dg} for West Africa. Meanwhile, Schuler \textit{et al.}\cite{Schuler:2001dg}, using a numerical weather prediction data, obtained a model which is more global in outlook relating  $T_m$  to $T_s$ as shown in Eq.$\;\ref{eq:5}$

\begin{equation}
\label{eq:5}
T_m=0.65T_s + 86.9
\end{equation}

From satellite data: which is of higher resolution spatio-temporally than the radiosonde data, it is the aim of this work to find a suitable statistical model for the evaluation of atmospheric mean temperature $T_m$ in West African region.


\section{Methodology} 
\subsection{Study Area} 
\begin{figure}[ht!]
	\centering
		\includegraphics[width = 9cm, height = 8cm]{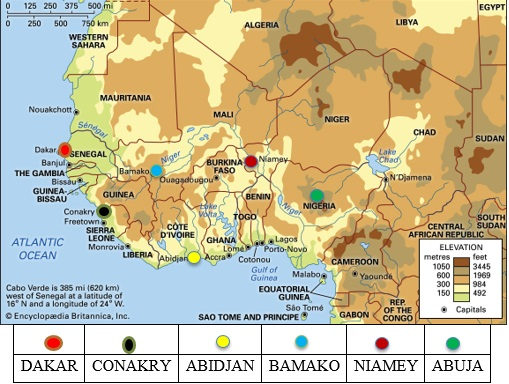}
	\caption{Map of West Africa showing Locations of study area (Encyclopeadia Britannica, $2018$).}
	\label{fig:Figure1}
\end{figure}

This study was conducted for locations in West African region as shown in Figure$\;\ref{fig:Figure1}$. The selected locations in West Africa for this research work are: Dakar ($14.76$\;$^oN$ Latitude, $17.36$\; $^oW$ Longitude), Conakry ($9.64\; ^oN$ Latitude, $13.57\; ^oW$ Longitude), Bamako ($12.63\; ^oN$ Latitude, $8.00\; ^oW$ Longitude), Abidjan ($5.36\; ^oN$ Latitude, $4.00\; ^oW$ Longitude), Niamey ($13.51\; ^oN$ Latitude, $12.12\; ^oE$ Longitude), and Abuja ($90.07\; ^oN$ Latitude, $7.39\; ^oE$ Longitude).

For ease of analysis of the data sets and climatic consideration, the study locations were further divided into ($1$) Hinterland, that is those locations in the interior of the study region and ($2$) Coastal, that is those regions close to the Atlantic Ocean. These divisions are shown in Table$\;$ \ref{tab:1}.

\begin{table*}[t]
\caption{Division of the Study Area into Hinterland and Coastal Regions.}
\label{tab:1}
\centering
\begin{tabular}{lllllr}
\toprule
\textbf{Station} &  & \textbf{Country} & \textbf{Longitude} & \textbf{Latitude} & \textbf{Period of data} \\
\midrule
\textbf{Dakar} & \multirow{3}{4em}{Coastal Region} & Senegal & $17.366\;^oW$ & $14.765\;^oN$ & $2004 - 2016$ \\
\textbf{Conakry} & & Guinea & $13.578\;^oW$ & $09.641\;^oN$ & $2004 - 2016$ \\
\textbf{Abidjan} & & Cote d'ivoire & $04.008\;^oW$ & $05.360\;^oN$ & $2004 - 2016$ \\
\midrule
\textbf{Bamako} &\multirow{3}{4em}{Hinter- land Region} & Mali & $08.003\;^oW$ & $12.639\;^oN$ & $2004 - 2016$ \\
\textbf{Niamey} & & Niger & $12.125\;^oE$ & $13.512\;^oN$ & $2004 - 2016$ \\
\textbf{Abuja} & & Nigeria & $07.399\;^oE$ & $09.077\;^oN$ & $2004 - 2016$ \\
\bottomrule
\end{tabular}
\end{table*}

\subsection{Data Analysis}
The data used in this work was obtained from the archive of the Climate Monitoring Satellite Application Facilities (CM-SAF) of the European Meteorological Satellites (EUMETSAT) at \href{https://www.cmsaf.eu/EN/Home/home{\_}node.html}{https://www.cmsaf.eu}. The data was obtained as a monthly average for a time period spanning from $2004$ to $2016$ in Network Common Data Format (netCDF).  The data obtained at six different pressure levels (level$1$ = $200$ $hPa$; level$2$ = $300$ $hPa$; level$3$ = $500$ $hPa$; level$4$ = $700$ $hPa$; level$5$ = $850$ $hPa$ and level$6$ (surface) = $1000$ $hPa$). are the specific humidity $q_m$ ($g/kg$) and the ambient temperature $T$ ($K$). The vapor pressure $e$ (in $hPa$) was estimated from the specific humidity using:

\begin{equation}
\label{eq:6}
e=\frac{0.5q_m}{10^{-3}q_m + 0.622}
\end{equation}

Panoply (v$4.3.1$), open source software from NOAA was used for the processing of the data while Jupyter notebook, an environment for running Python code, was used for data plotting, curve fittings and other statistical analysis.

Goodness of fit and correlational analysis of the derived empirical equations were done using coefficient of determination ($R^2$), mean-biased error ($MBE$) and root-mean-square error ($RMSE$).


\section{Results and Discussion}
The derived empirical equations relating the mean atmospheric temperature $T_m$ to the surface temperature $T_s$ for each of the locations considered in this work within the West African region are shown in Table$\;\ref{tab:2}$. All the stations considered show good linear correlations between $T_m$ and $T_s$Except for Conakry with a below average coefficient of determination. It will be observed from Table$\;\ref{tab:2}$ that the coastal region generally does not show good linear correlation between $T_m$ and $T_s$, whereas the hinterland region shows very good linear correlation: also it was observed that the ambient surface temperature at the coast is generally much lower than that at the hinterland of West Africa.
\begin{table*}[t]
\caption{Statistics for each of the stations in West Africa.}
\label{tab:2}
\centering
\begin{tabular}{llccc}
\toprule
\textbf{Location}& & \multicolumn{2}{c}{\thead{\textbf{Empirical} \\\textbf{Model}\\$T_m = aT_s + b$}} & \thead{\textbf{Coefficient} \\ \textbf{of}\\ \textbf{Determination}} \\
\cmidrule(r){3-4}
\textbf{Name} & & \textbf{$a$} & \textbf{$b$} & \textbf{$R^2$} \\
\midrule
Dakar & \multirow{3}{4em}{Coastal Region} & $0.57$ & $123.15$ & $0.575$ \\
Conakry & & $1.19$ & $63.28$ & $0.442$ \\
Abidjan & & $0.87$ & $32.68$ & $0.741$ \\
\midrule
Bamako & \multirow{3}{4em}{Hinter- land Region} & $0.68$ & $89.96$ & $0.646$ \\
Niamey & & $0.81$ & $52.23$ & $0.763$ \\
Abuja & & $0.85$ & $40.88$ & $0.902$ \\
\bottomrule
\end{tabular}
\end{table*}

\begin{figure}[ht!]
  \includegraphics[width= 10cm, height = 9cm]{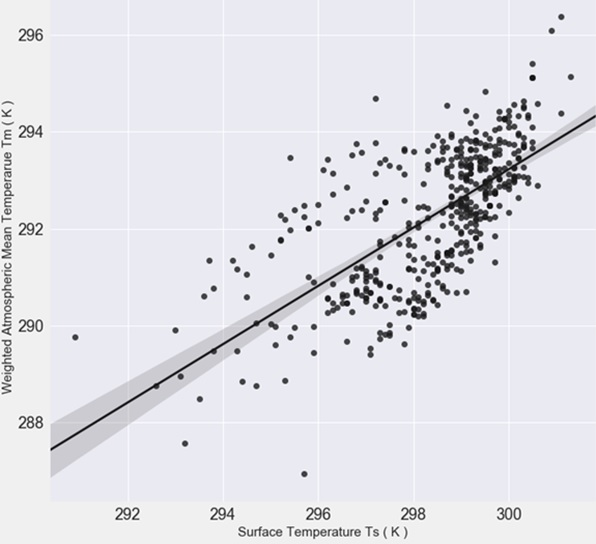}
  \caption{Coastal Region data plot and linear fit.}
  \label{fig:2}
\end{figure}

\begin{figure}[ht!]
  \includegraphics[width= 10cm, height = 9cm]{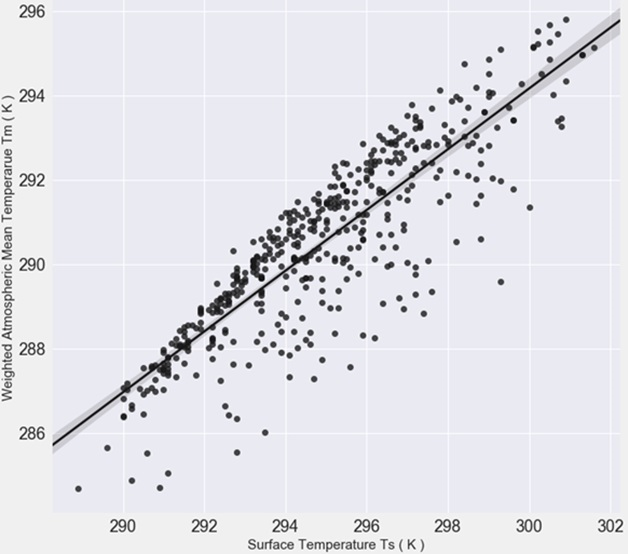}
  \caption{Hinterland Region data plot and linear fit.}
  \label{fig:3}
\end{figure}

Figures$\;\ref{fig:2}$ and $\;\ref{fig:3}$ show the combined plot and linear fit to the dataset of coastal and hinterland respectively. For these figures empirical equations Eq.$\;\ref{eq:7}$ and Eq.$\;\ref{eq:8}$ were derived for the coastal and hinterland respectively. It will be observed again from Figures$\;\ref{fig:2}$ and$\;\ref{fig:3}$ as well as the derived Eqs.$\;\ref{eq:7}$ and $\;\ref{eq:8}$ that the model fit to the hinterland data has better correlation than the fit to the coastal data.

\begin{equation}
\label{eq:7}
T_m=0.60T_s + 112.73 \; \; \; \; \; R^2 = 0.459
\end{equation}

\begin{equation}
\label{eq:8}
T_m=0.72T_s + 77.89 \; \; \; \; \; R^2 = 0.752
\end{equation}

\begin{figure}[ht!]
  \includegraphics[width= 10cm, height = 9cm]{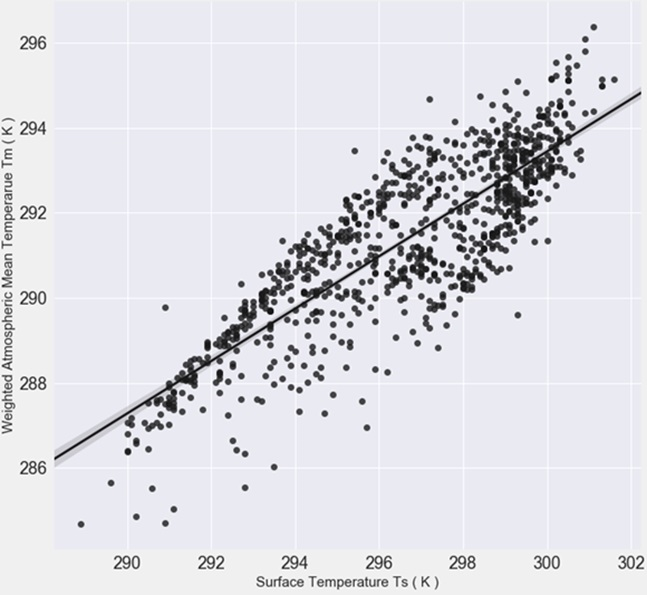}
  \caption{A combined West African data plot and linear fit.}
  \label{fig:4}
\end{figure}

Figure$\;\ref{fig:4}$ shows a plot of all the combined data for West Africa. The statistical model derived from Figure$\;\ref{fig:4}$ is given in Eq.$\;\ref{eq:9}$ with the corresponding coefficient of determination:

\begin{equation}
\label{eq:9}
T_m=0.617T_s + 108.49 \; \; \; \; \; R^2 = 0.709
\end{equation}

Isioye \textit{et. al.}\cite{Isioye:2016dg} using radiosonde data covering $2009$ to $2013$ for West Africa obtained a corresponding statistical model as given in Eq.$\;\ref{eq:10}$:

\begin{equation}
\label{eq:10}
T_m=0.5743T_s + 116.60 \; \; \; \; \; R^2 = 0.436
\end{equation}

It will be noted that Eq.$\;\ref{eq:9}$ shows a better statistical performance than Eq.$\;\ref{eq:10}$, though the two statistical models could be observed to be quite similar. A statistical comparison of Eq.$\;\ref{eq:9}$ with Eq.$\;\ref{eq:10}$ yields root-mean-square error ($RMSE$) of $1.189$ and mean-biased error ($MBE$) of $0.0953$, further showing that the two statistical models are comparable. Satellite data has wider spatio-temporal coverage than radiosonde data hence, the better coefficient of determination $R^2$, observed for Eq.$\;\ref{eq:9}$ as compared to Eq.$\;\ref{eq:10}$.

\section{Conclusion}
Using satellite data from the Satellite Application Facility (CM-SAF) of the EUMETSAT a suitable statistical model has been derived to estimate weighted atmospheric mean temperature $T_m$. The statistical model derived is a simple linear model of $T_m$ as a function of surface temperature $T_s$.

The derived model was compared with a similar statistical model which had earlier been derived from the radiosonde data in West Africa. The correspondence between the two models were found to be significant.

\textbf{Acknowledgment}\newline
This work was done using data from EUMETSAT’s Satellite Application Facility on Climate Monitoring (CM SAF). DOI: $10.5676$/EUM{\_}SAF{\_}CM/WVT{\_}ATOVS/V$001$. 


\end{document}